# The 6-Ds of Creating AI-Enabled Systems

**Dr. John Piorkowski**

Johns Hopkins University Applied Physics Laboratory
11100 Johns Hopkins Rd. Laurel, Maryland 20723-6099
John.Piorkowski@jhuapl.edu

## Abstract

We are entering our tenth year of the current Artificial Intelligence (AI) spring, and, as with previous AI hype cycles, the threat of an AI winter looms. AI winters occurred because of ineffective approaches towards navigating the technology "valley of death." The 6-D framework provides an end-to-end framework to successfully navigate this challenge. The 6-D framework starts with problem decomposition to identify potential AI solutions, and ends with considerations for deployment of AI-enabled systems. Each component of the 6-D framework and a precision medicine use case is described in this paper.

## 1. Introduction

In 2012, a team of researchers, led by Geoffrey Hinton, advanced the field of computer vision using convolutional neural networks during the ImageNet competition (Krizhevsky et al., 2012). This significant machine learning result fueled the current AI spring, now approaching a decade of existence. Of course, history often repeats itself, so is another AI winter looming? Looking at the cycle of AI springs and winters, winters have occurred partly because of ineffective transiting the technology "valley of death."

One technique for successfully navigating this challenge is to approach technology solutions with an end-to-end engineering perspective. This paper describes an end-to-end framework for creating an AI-enabled system. The concept of the 6-D framework is an extension of Neil Lawrence's paper (Lawrence, 2019) on the 3-Ds of machine learning. Furthermore, the 6-D framework is influenced by experience across a set of domains focused on applied research. The 6-D framework starts with the initial ideation of AI-enabled systems and continues through design, diagnosis, and deployment.

## 2. Overview of the 6-D Framework

The 6-D framework (Figure 1) takes a holistic view of creating AI-enabled systems from the initial concept through design and deployment. In an AI-enabled system, an AI algorithm does not stand alone and is part of a more extensive system or capability to achieve broader business or mission objectives. It is important to note that ethics, which is a critical consideration in many AI applications, is cross-cutting and deserves attention in each component. The following sections describe each component in greater detail.

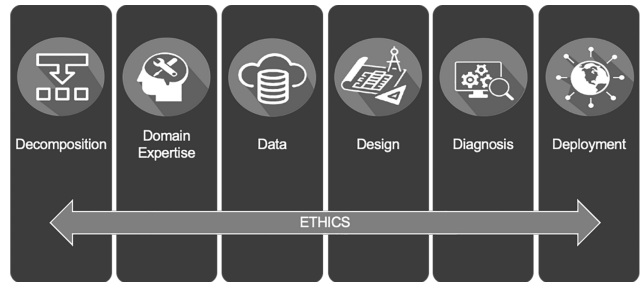

### 2.1 Decomposition

As with many over-hyped technologies, AI technology is often pursued without significant thought on the underlying problem that could benefit from AI or on the question of whether AI technology is the best option. This is one reason that many technologies do not transit the technology valley of death. Elwood et al., 2020, recommended innovation processes for addressing this challenge.

The innovation process recommendations include: refining the narrative of the technology concept, understanding how technology will be used, conducting a technical evaluation of lab models and comparative value assessment, and integrating inputs from innovation actors, i.e., key stake-holders seeking some future end state. Each of these processes involves some appreciation of user value in the technology.

When creating an AI-enabled system, a developer can use the decomposition component of the 6 D framework to address refining the technology concept, understanding technology use, and assessing value.

During decomposition, a developer can employ various techniques to frame the problem the AI enabled system is seeking to solve. While there is no single approach, the following four techniques can be adopted to perform a rigorous decomposition:

1. Envisioned Futures
2. System Decomposition
3. Reinventing Jobs
4. Design Thinking

Envisioned Futures (Collins and Porras, 1996) allows organizations to build a company's vision by asserting a far-reaching Big Hairy Audacious Goal (BHAG) coupled with vivid description. The vivid description entails creating a specific, concrete view of the future when the BHAG is achieved. Creating vivid descriptions often uncovers the need for AI solutions in modern business settings. Well-written vivid descriptions can offer new insights into motivating AI technologies that can reduce human labor or speed up processes.

A second technique, system decomposition, comes from the classic systems engineering field. Systems decomposition is part of the systems development lifecycle, represented as a V (Ravitz et al., 2013). The systems development lifecycle begins with requirements elicitation through engagement with stakeholders. The second part of system development is requirements documentation.

A third technique, Reinventing Jobs, (Jesuthasan and Boudreau, 2018) describes a four-step process for automating work: deconstructing the job, assessing its strategic value, identifying automation options, and optimizing the outcome. An appropriate step in this process is considering the strategic importance of the deconstructed jobs. Applying an AI solution can be resource intensive, so a strategic assessment is critical.

Finally, a popular technique for decomposing a problem to discover opportunities for AI applications is using design-thinking methods. Chasanidou et al. (2015) provided a survey of these methods. They describe the general design-thinking process, which includes empathizing, defining, ideating, prototyping, and testing. The stages of empathizing and ideating involve engaging with stakeholders and brainstorming with diverse stakeholders.

## 2.2 Domain Expertise

As noted in the decomposition section, AI technology can help improve a situation in terms of reducing workload, speeding up a process, or achieving a new insight. However, understanding the problem in terms of business objectives or operations requires individuals with domain knowledge. Participation from domain experts who also serve as innovation actors increases the likelihood of an AI project succeeding, because domain experts advocate for change. As evidenced from creating AI-enabled systems in healthcare, disaster response, and national security, the role of domain experts is critical to the success of an AI project. They play an important role beyond decomposition, providing key support during the subsequent phases of the framework. For example, domain experts bring credibility to community members who will be the ultimate users of AI-enabled systems, and can assist with adoption of new technology.

## 2.3 Data

In AI projects, the data engineering phase is generally the most resource intensive, involving collecting, moving, storing, transforming, labeling, and optimizing data to facilitate the design of the system. The data effort stems from the fact that modern AI-enabled systems will include elements of machine learning. Yuji Roh et al., 2021, provide a survey of the data collection landscape for machine learning. Their work recognizes the convergence of machine learning and data management research fields, including data analytics and data science.

Another data engineering perspective is the Extract, Transform, and Load (ETL) process that is referenced in the context of the data lifecycle (Gavin, 2019). The extraction step addresses the collection of raw data using a range of various methods and quality. The transformation is critical and generally is the most resource intensive step. A 2018 Forbes survey of data scientists indicated that 80% of their time is spent on data preparation.

In machine learning, the data must be engineered to allow feature representations. Neil Lawrence offers a framework to assess data readiness for analytics (Lawrence, 2017), using three levels. The lowest level (C-Level) describes the challenges with data engineering and wrangling. As Lawrence explains, many organizations claim they have data, but the data have not been made available for analytic use. He refers to this type of data as "hearsay" data. B-Level data require an understanding of its faithfulness and representation. Finally, A-Level represents data in context, assessed to be able to answer organizational questions.

Another key AI technique is embedding spaces[1] that represent high-dimensional vector spaces in low-dimensional vector spaces. An embedding[2] is a translation of a high-dimensional vector into a low-dimensional space. Embedding techniques vary by the type of data. The techniques and data engineering methods are associated with the data types. Data used to create AI-enabled systems vary, and are shown in the categories described in Table 1.

Table 1. Data Categories and Engineering Processes

| Data Category | Data Engineering Processes |
|---|---|
| Computer Vision | Image manipulation<br>Filtering<br>Image transformations |
| Unstructured Text | Tokenization<br>Stemming<br>Lemmatization<br>Stop words removal<br>Embedding |
| Categorical | One hot coding<br>Imputation |
| Time Series | Time alignment<br>Feature generation |

The second aspect of data is management and storage. The current AI spring renaissance is fueled by data availability. Developers have several alternatives for organizing and storing data in AI enabled systems.

Historically, relational database systems became the most prevalent data storage technique in business operations starting in the 1980s. A relational database involves creating tables and using keys for relationships between tables. The advantage of relational databases is that they are efficient at minimizing data redundancy and maximizing data integrity. One disadvantage of relational databases is their inability to deal with large amount of unstructured data, varying data types, and a large number of users querying.

In the late 2000s, we entered the "big data era" and the dawn of No-SQL or non-relational data storage approaches. These approaches were motivated by exponential growth in data in the internet. The family of No-SQL approaches include:

- key-value pair databases;
- document databases (JSON or XML);
- column family store databases; and
- graph databases.

Finally, practical AI-enabled systems often employ "polyglot persistence" that entails a mix of relational and non-relational data storage approaches.

## 2.4 Design

The AI field is characterized by several algorithmic approaches ranging from rule-based systems to machine learning algorithms. A simple framework for designing AI-enabled functions comprises perceive, decide, or act. Table 2 provides a high-level framework for mapping various algorithm classes into the three functions. Through the decomposition step, a problem is examined to identify potential applications for AI technologies. In practice, many AI-enabled solutions may include a combination of algorithm classes.

The data section reviewed various data categories and engineering techniques. The algorithmic techniques in Table 2 can be applied to various data categories. Initially, developers emphasized computer vision data in deep neural networks for supervised learning. With the expert level performance achieved with deep neural networks in computer vision data, developers have since applied them to other data types such as text and time series.

Table 2. High-Level Framework Mapping of AI Algorithm Classes

| Algorithm Class | Algorithm Subclass | Perceive | Decide | Act |
|---|---|---|---|---|
| Supervised ML | • Regression<br>• Traditional Classification<br>• Deep Neural Networks | X | | |
| Unsupervised MS | • Clustering<br>• Generative Adversarial Networks | X | | X |
| Reinforcement Learning | | | X | |
| Rules and Logic | | | X | X |

## 2.5 Diagnosis

The diagnosis component addresses how AI-enabled systems are assessed and what metrics are used. The design section identified four algorithm classes that bring a different set of metrics. Typical metrics for supervised machine learning algorithms include accuracy, a confusion matrix, per class accuracy, log loss, precision, recall, mean average pre-cision, and Area Under the Curve (AUC). Metrics for unsupervised algorithms are more complicated because ground truth data may not exist to apply metrics such as accuracy. Since unsupervised learning involves clustering, metrics exist to measure the cohesion of clusters.

---

[1] https://www.forbes.com/sites/gilpress/2016/03/23/data-preparation-most-time-consuming-least-enjoyable-data-science-task-survey-says/?sh=638c993a6f63

[2] https://cloud.google.com/architecture/overview-extracting-and-serving-feature-embeddings-for-machine-learning

Reinforcement Learning systems are evaluated with a different set of metrics. These metrics are summarized by Henderson et al., 2018 and include confidence bounds, power analysis, and significance of the optimization.

## 2.6 Deployment

There is no one-size-fits-all approach for AI-enabled system deployment, but it is important to note that AI technologies are generally employed in a broader system context. With the prevalence of cloud computing and client/server models, AI-enabled systems leverage this type of computing paradigm. For AI-enabled systems that employ client-server models in a cloud environment, computation for AI models resides on the server side. Alternatively, the AI models can be hosted on local devices. Factors to consider in selecting a deployment model include governance of data and models, computational requirements, and connectivity requirements. Another tradeoff for deployment includes whether to use Central Processing Units (CPUs) or Graphical Processing Units (GPUs). GPUs are heavily leveraged in deep neural networks given their advantages with parallel processing of vectorized data. CPUs are lower cost and have advantages for applications that require significant computer memory.

Hidden technical debt and continuous monitoring are other vital factors to consider for deploying AI-enabled systems. Sculley et al., 2015 used the software engineering concept of technical debt to describe significant ongoing maintenance costs in real-world machine learning systems. A key point in the document is that the algorithm portion of the software constitutes a small percentage of the overall code base of an AI-enabled system.

Continuous monitoring addresses changes in the behavior of deployed AI models. One common problem when supervised machine learning systems are deployed is concept drift, resulting when the distribution of data fed into the model deviates from the distribution of training data used to create the model. Gama et al., 2014 provide a survey of methods for concept drift mitigation.

## 3. Precision Medicine Use Case

Piorkowski, 2021 describes several domains where data science and AI have been applied. One example includes the Precision Medicine Analytical Platform (PMAP) that enables AI for precision medicine. The PMAP design commenced in 2016 with decomposition of the precision medicine domain, initially focused on prostate cancer and multiple sclerosis.

The decomposition phase led to challenges that can be solved by the application of AI to treat these diseases. This phase involved close collaboration with domain experts who are renowned in the treatment of prostate cancer and multiple sclerosis.

The close collaboration with domain experts and designers helped framed subsequent data engineering and design phases. Data engineering started with curating medical record data in EPIC, together with other data available in diagnostic imaging systems (e.g., CT Scans).

The data engineering involved handling all types of data mentioned in the previous data section, and represented the most resource-intensive step. With the data engineering step completed, new AI solutions were created for clinicians. For prostate cancer, a new capability was designed using Natural Language Processing (NLP) for machines to discover patient Gleason Scores in medical records (Chee et al., 2018). Special efforts to include independent human annotations were used to diagnosis the NLP capability.

Finally, a deployment system was created in a Microsoft Azure cloud environment that provides governance mechanism for medical data. Additional tools are included such as python and R for clinicians seeking to do research.

## 4. Summary

This paper introduces the 6-D framework for creating AI-enabled systems. The 6-D framework provides an end-to-end holistic list of critical elements that should be addressed to build AI capabilities successfully. The topic of ethics in AI systems is recognized as cross-cutting, and ethical considerations should be addressed as part of each element of the 6-D framework.

## 5. Abbreviations and Acronyms

| | |
|---|---|
| AI | Artificial Intelligence |
| AUC | Area Under the Curve |
| BHAG | Big Hairy Audacious Goal |
| CPU | Central Processing Units |
| ETL | Extract, Transform, and Load |
| GPU | Graphical Processing Units |
| NLP | Natural Language Processing |
| PMAP | Precision Medicine Analytical Platform |